\documentclass[global, twocolumn]{svjour}
\usepackage{latexsym}
\usepackage{graphics}
\usepackage[draft]{hyperref}
\usepackage{amsmath}
\usepackage[usenames,dvipsnames]{color}
\usepackage{ulem}
\journalname{}
\begin{document}
\title{Demonstration of a Transportable 1~Hz-Linewidth Laser}
\author{Stefan Vogt\inst{1}, Christian Lisdat\inst{1}, Thomas Legero\inst{1}, Uwe Sterr\inst{1}, Ingo Ernsting\inst{2}, Alexander Nevsky\inst{2}, Stephan Schiller\inst{2}}
%\author{First author\inst{1} \and Second author\inst{2}% etc
%}                     % Do not remove

%
\institute{Physikalisch-Technische Bundesanstalt (PTB), Bundesallee 100, 38116 Braunschweig, Germany
\and Institut f\"ur Experimentalphysik, Heinrich-Heine-Universit\"at D\"usseldorf, 40225 D\"usseldorf, Germany\\$^*$Corresponding author: christian.lisdat@ptb.de, step.schiller@uni-duesseldorf.de}
\date{Received: date / Revised version: date}
\authorrunning{Stefan Vogt et al.}
\maketitle
\begin{abstract}
We present the setup and test of a transportable clock laser at 698~nm for a strontium lattice clock. 
A master-slave diode laser system is stabilized to a rigidly mounted optical reference cavity. 
The setup was transported by truck over 400~km from Braunschweig to D\"usseldorf, 
where the cavity-stabilized laser was compared to a stationary clock laser for the interrogation of ytterbium (578~nm). Only minor realignments were necessary after the transport. 
The lasers were compared using a Ti:Sapphire frequency comb as a transfer oscillator. The generated virtual beat showed a combined linewidth below 1~Hz (at 1156~nm). The transport back to Braunschweig did not degrade the laser performance, as was shown by interrogating the strontium clock transition.
\end{abstract}%
%\subsection*{Introduction}
Optical clocks based on trapped cold atoms are now outperforming the best microwave clocks, thus enabling new studies and applications.
Operated in space and on the ground at different locations, they could enable relativistic geodesy and improved fundamental physics tests \cite{sch09,wol09}.
%, and on the ground they could be operated  comparison of clocks at remote sites, 
Reliable and rugged optical clocks with high stability and accuracy are therefore an important need. So far, high-performance cold atom optical clocks are still bulky laboratory setups that cannot easily be transported. Therefore developments towards transportable optical clocks are required and have been initiated \cite{sch09b,sch10d,lei11}.
\begin{figure}[htb]
\resizebox{\columnwidth}{!}{
\includegraphics{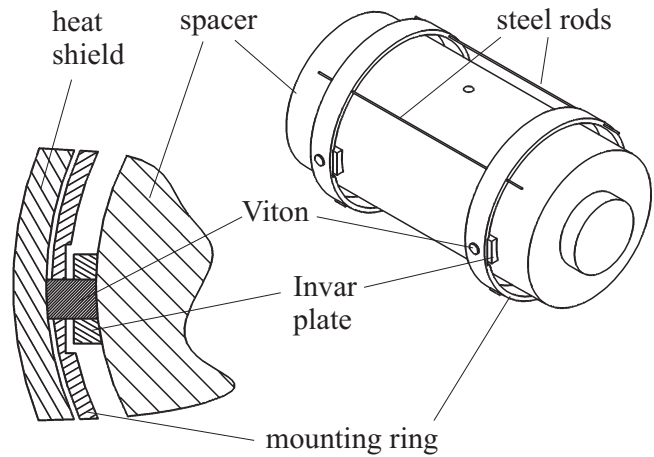}
}
\caption{Sketch of the semi-rigid mounting of the transportable reference resonator.}
\label{fig:cavity}
\end{figure}
One of the most critical parts concerning transportability is the optical reference cavity that is used as a flywheel to ensure a high short-term frequency stability of the interrogation laser between the atom interrogation cycles.
To achieve an optical linewidth of the clock laser in the range of one Hertz, the cavity has to be isolated from all external disturbances. Usually, the cavity is mounted very loosely in a vibration insensitive configuration \cite{not05,naz06,web07} to avoid excessive forces that would lead to deformations of the cavity and fluctuations of the optical path length and therefore its  resonance frequency.
However, this prevents the cavity from being easily transported, because the resonator will move and might be damaged during transportation.

In this letter, we describe a prototype clock laser for a transportable neutral atom lattice clock. As a first realistic test for future transportable clock setups it was transported from Braunschweig to D\"usseldorf.
%A schematic of the setup when the clock laser comparison was performed is shown in Fig.\,1.

%\subsection*{Transportable Laser Setup}

The design of the laser system is constrained by requirements for its transportability in a car or truck and its spectral purity, for which typically a linewidth of 1~Hz and a fractional stability of $10^{-15}$ on timescales of a few seconds is needed.

The transportable clock laser setup consists of three separate parts: the laser breadboard (60~cm $\times$ 90~cm), a rolling table with the reference cavity inside a box for acoustic noise isolation (outer dimensions including the table 150~cm $\times$ 88~cm $\times$ 195~cm) and an electronics rack (60~cm $\times$ 60~cm $\times$ 180~cm). No effort was made to compactify the electronics. The laser breadboard includes a master-slave setup with a master laser in Littman configuration. The laser is locked to the reference cavity with the Pound-Drever-Hall technique using the laser current for fast frequency control \cite{leg09}. Standard optical components were used to offer all the functionality that is needed for clock laser operation such as shifting the laser frequency with acousto-optic modulators (AOM). In particular the path length stabilization \cite{ma94,wil08a} for two fibers was also set up using AOMs on the breadboard. The light from the laser was sent to the cavity via a 2~m long fiber which was not actively stabilized. The volume of the system without the electronics rack was less than 3.0~m$^3$. For transportation the laser breadboard was packaged in a wooden box, the optical fibers and most of the electrical connections were removed. No special care was taken to isolate the laser system or the cavity against vibrations while transporting it.

The cavity is mounted in a passive, gold plated copper heat shield inside a vacuum chamber, evacuated by a small 2~l/s ion getter pump. The vacuum system is installed on a passive vibration isolation platform, which was locked during the transport. The temperature of an aluminum tube enclosing the vacuum system was stabilized to $24\,^{\circ}$C using a heating foil as control element. For simplicity and to avoid possible condensation no cooling was employed.
This temperature does not agree to the zero expansion temperature of the all-ULE (ultra-low expansion glass) cavity. From measurements done at similar spacers made of the same ULE batch we expect this temperature to be at about $11\,^{\circ}$C. With a slope of the thermal expansion coefficient of $2 \times 10^{-9}$~K$^{-2}$ we expect at $24\,^{\circ}$C a sensitivity of the resonator frequency to temperature fluctuations of approximately $-10$~kHz/mK. During the transport, an uninterruptible power supply unit was used to keep the the vacuum pump and the temperature stabilization of the cavity running.

The cylindrical cavity spacer has a diameter of $50$~mm  and is $100$~mm long. One plane and one concave (1~m radius of curvature) ULE mirror (diameter 25.4~mm, thickness 6.3~mm) were optically contacted to the spacer. The finesse of the cavity is 330~000.
To ensure transportability the mounting of the reference resonator was designed to be much more rigid than for other ultra-narrow linewidth lasers \cite{web07,naz06}. Small Viton cylinders are tightly pressed into four holed Invar plates glued onto the cavity spacer (Fig.~\ref{fig:cavity}, details given in \cite{leg09}). The cylinders (and thus the cavity) are held in mounting rings that fix the resonator position in all three directions.  Side-way movement is inhibited because the Viton pieces tightly fit into holes of the mounting rings and touch the inner side of the heat shield. The suspension positions of the mounting points were optimized by finite element (FEM) calculations. The top of the cylinders is located at the horizontal symmetry plane and 17.5~mm from the end faces of the spacer. In previous measurements \cite{leg09} large frequency changes of the resonator mode were observed under accelerations. It was assumed that unintended mechanical contact of the spacer to the housing was the source of this effect.
% rectifying this issue indeed resulted in a lower sensitivity to vertical vibrations of 120~kHz/ms$^{-2}$; however it  is still significantly larger than the expected value. 

In a refined layout the mounting rings were connected by steel rods and great care was employed to prevent the cavity from touching the lids of the surrounding heat shield. Also the length of the Viton cylinders was increased to allow for a tight reliable fit of the cavity. However only a marginally lower sensitivity to vertical vibrations of $\Delta L / L \approx 2.7 \times 10 ^{-9}$/g was achieved. We now attribute this effect to squeezing forces introduced by the cavity mounting. These forces arise because the Viton pieces are squeezed between the heat shield and the cavity spacer. For a force of 1\,N on each cylinder we calculated a change of the cavity length $\Delta L / L \approx 1 \times 10 ^{-8}$ if the force acts in axial direction or $\Delta L / L \approx 3 \times 10 ^{-9}$ if the force is radially squeezing the spacer. Under acceleration a reaction force of 4~N/g has to be supplied by the Viton cylinders. A 30\% coupling of this force into axial direction could explain the observed sensitivity. Also deformations of the heat shield under acceleration can change the squeezing forces and contribute to the vibration sensitivity.

Meticulous elimination of vibration incoupling on the vibration isolation platform led to acceptable laser performance. We also used the virtual beat with an ultra-stable laser at 657~nm \cite{naz06} to optimize our laser setup. A width of the beat signal of about 1~Hz was achieved. Calculating the Allan deviation from the counted beat note we observed a flicker floor at about $2 \times 10^{-15}$ fractional stability.

%\subsection*{D\"usseldorf Setup}
The clock laser for the D{\"u}sseldorf ytterbium lattice clock is based on a 1156~nm diode laser \cite{Nev08} with an external Littrow configuration cavity and containing a home-made intracavity electro-optic phase modulator for fast frequency corrections. 
The free-running linewidth is less than 100~kHz. 
The output radiation (approx. 20~mW) is focused into a 20~mm PPLN waveguide for frequency doubling. %(at \mbox{T = $14.6\,^{\circ}$C)} by a $f = 3$~mm aspheric lens.
A maximum output of 0.5~mW at 578~nm was produced. %This wave is sent into a ULE cavity. A standard Pound-Drever-Hall technique is used for generating an error signal. Phase modulation at \mbox{12 MHz} was provided by an external EOM, and a servo acts on the intracavity modulator for fast control and on the laser current for slow frequency control.
This wave is stabilized to an all-ULE cavity with a standard Pound-Drever-Hall scheme.
The finesse of the cavity exceeds 300~000. It is of the cut-out type \cite{web07}, with 100~mm length, 50~mm diameter, a 30~mm deep cut-out starting 4~mm below the mid-plane. %The mirror substrates are of ULE glass %and coated for \mbox{578 nm}. 
%The cavity is housed  in a gold-plated copper box and both are contained in a compact (11~cm $\times$ 11~cm $\times$ 18~cm) aluminum vacuum chamber.  Both the vacuum chamber and the copper housing are actively temperature stabilized by thermoelectric cooling devices.
\begin{figure} [ht]
\resizebox{\columnwidth}{!}{%
  \includegraphics{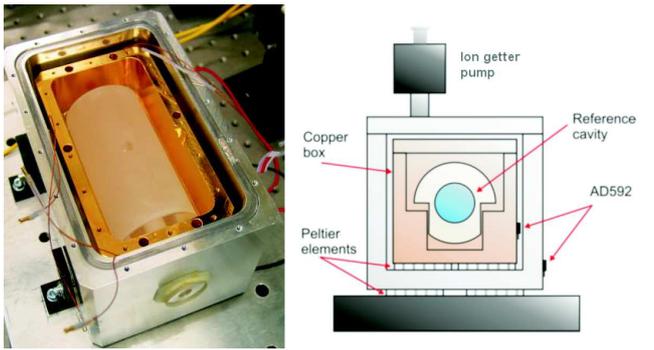}
}
\caption{Left: photograph of the reference cavity inside the compact vacuum chamber with removed cover and insulation. Right: schematic of the vacuum chamber, showing the elements of the two-stage temperature stabilization system, with AD592 temperature sensors (insulation box is not shown)  }
\label{fig:DDCavity} 
\end{figure}

The 578~nm reference cavity is housed in a massive gold-plated copper box and both are contained in a compact (11~cm $\times$ 11~cm $\times$ 18~cm) aluminum vacuum chamber (see Fig. \ref{fig:DDCavity}). The two entrance windows are glued to the vacuum chamber using Torr Seal epoxy. The chamber is machined out of a single aluminum block; the cover is attached to the chamber using  M4 screws and indium foil for sealing. A vacuum of about 5 $\times 10^{-8}$~mbar is maintained using a 2~l/s ion getter vacuum pump, attached to the cover plate. The vacuum chamber and the vacuum pump are covered with a thermal and acoustic insulation box.

Both the vacuum chamber and the copper housing are independently actively temperature stabilized by Peltier elements and two AD590 temperature sensors. With this two-stage temperature stabilization system, which includes a temperature stabilized electronic controller, we reach a temperature instability of 0.15~mK at 100~s and 0.07~mK at 10~000~s integration time, as measured by the in-loop sensor on the copper box.

The temperature of the cavity is kept close to the ULE's thermal expansion coefficient zero crossing at approx. $20\,^{\circ}$C. The described construction demonstrated a remarkable performance as measured with the frequency comb in comparison to a GPS disciplined hydrogen maser. No influence of the residual temperature instability on the resonator frequency could be detected within the  stability of the maser (approx $10^{-12}$). Over a period of 35 days, the average drift of the cavity frequency was 56~mHz/s.

All the optics hardware fits on a $90\,$cm $\times$ $90$~cm breadboard that is supported by an active vibration isolation system, providing isolation along the three spatial directions. Except for lock electronics, the system is enclosed by a wooden box for acoustic protection. This clock laser had previously been characterized by comparison with a second, independent ULE cavity. % for \mbox{578 nm}. 
A beat with an instability of less than $2.5\times10^{-15}$ for 1~s -- 10~s was obtained (after subtraction of the differential cavity  drift of approx. 0.1~Hz/s).
\begin{figure}[ht]
\resizebox{\columnwidth}{!}{%
  \includegraphics{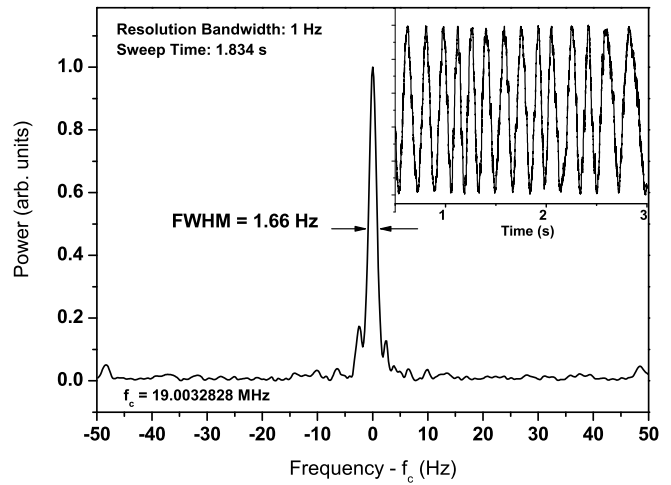}
}
\caption{Virtual beat at 259~THz (1156~nm) monitored with a spectrum analyzer. The inset shows 2.5 seconds of the digitized virtual beat signal mixed down to a few Hertz. The two measurements were taken at slightly different times.}
\label{fig:beat}   
\end{figure}

%\subsection*{Results}
\begin{figure} [ht]
\resizebox{\columnwidth}{!}{%
  \includegraphics{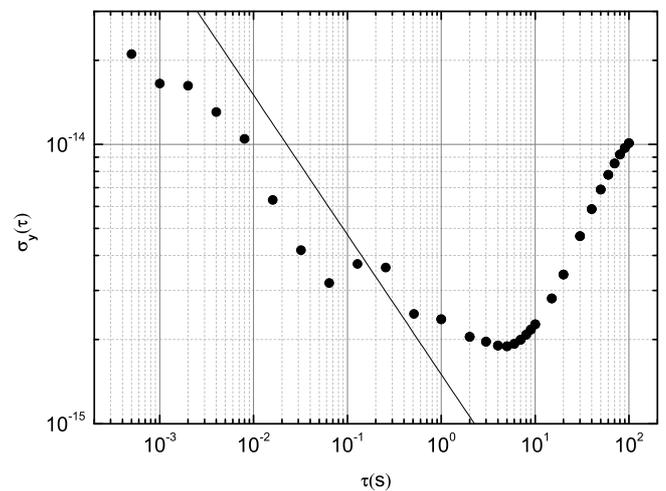}
}
\caption{Fractional Allan deviation of the virtual beat between the two clock lasers during a quiet period. For times shorter than one second the values where derived using the Hilbert transform of the mixed down beat signal (see also text). For longer times we used a frequency counter. Linear and quadratic cavity drifts were removed in both cases. The straight line corresponds to the stability of a virtual beat at 1156~nm with white frequency noise and 1~Hz width.}
\label{fig:allan} 
\end{figure}

The 5 hour transport of the 698 nm laser system was performed using a small transport truck. % with air dampers
The Sr laser was locked to its ULE cavity in D\"usseldorf within one day after start of loading in Braunschweig. Some realignment of the optical path had to be done after the transport because two mirror holders were not fixed well enough, but there was no indication that the cavity had moved inside its mounting.
For the characterization of the Sr clock laser the two lasers were operated in the frequency comb laboratory room in D\"usseldorf. Both laser beams were sent via fibers to the frequency comb, the fiber noises were actively canceled.

The optical frequency synthesizer is a commercial Ti:Sa-based frequency comb (Menlo Systems), modified in-house for higher long-term stability and expanded emission range. The 698~nm and 1156~nm laser beams (frequencies $\nu_{698},\nu_{1156}$) are individually superposed with the comb via two beat lines. In this beat lines the laser beams are coupled out of their fibers and superposed with the corresponding modes from the comb. Appropriate band-pass filters block the unnecessary comb modes before the overlap. The generated beat notes are detected by two photodetectors. The beat notes with frequencies $\Delta_{698}$, $\Delta_{1156}$ between each laser wave and the nearest comb mode were filtered and amplified with two tracking oscillators. The frequency comb's repetition rate and carrier envelope frequency $f_\mathrm{CEO}$ were locked to a GPS disciplined hydrogen maser.
Counting $\Delta_{1156}$ and $\Delta_{698}$ enabled absolute frequency measurement of the two clock lasers to be performed simultaneously.

The virtual beat technique \cite{tel02b} consists in generating by electronic means an r.f. signal oscillating at the frequency
$\Delta_{vb}=(f_\mathrm{CEO}+\Delta_{1156}) - (m_1/m_2)(f_\mathrm{CEO}+\Delta_{698}),$
that shows the properties of a beat note between two lasers at 1156~nm, since $\Delta_{vb}=\nu_{1156} - (m_1/m_2)\nu_{698}$. Here $m_1$ ($m_2$) is the mode number of the nearest comb line to the 1156 nm (698~nm) clock laser radiation.
This beat note is sent to a spectrum analyzer, 
counted with a dead-time-free counter and, after mixing it down to a few Hertz, sampled by an analog-digital converter. A record of the spectrum analyzer is shown in Fig.~\ref{fig:beat} together with a short sample of the digitalized beat (inset).

The fractional Allan deviation shown in Fig.~\ref{fig:allan} was calculated using two different methods. For time values larger than 1\,s we used the counter signal, while the shorter values were derived from the signal of the analog-digital converter. For this purpose we used a Hilbert transform to add an imaginary part to the real signal to obtain an analytic signal \cite{Shm06}. From this signal the phase and the instantaneous frequency were extracted. After removing linear and quadratic drifts the Allan deviation for time intervals as short as $0.5$\,ms was calculated.
For short times the trend is consistent with the linewidth observed in Fig.~\ref{fig:beat}. From 1~s to 10~s a noise floor of $2 \times 10^{-15}$ is reached. Beyond 10~s, the residual differential cavity drift dominates the stability.
We conclude from Allan deviation and virtual beat spectrum that both lasers have a linewidth close to 1~Hz.%, which is an exceptional result for a transportable laser system.

\begin{figure} [htb]
\resizebox{\columnwidth}{!}{%
  \includegraphics{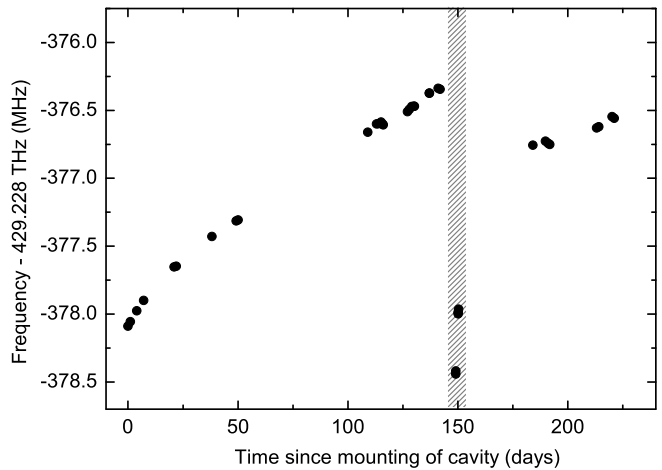}
}
\caption{Frequency of one cavity mode plotted over 8 months. The hatched area shows the measurements during the one week stay in D\"usseldorf.}
\label{fig:drift}      
\end{figure}

We have monitored the resonance frequency of the transportable cavity over more than 200 days including the time in D\"usseldorf. A shift of about 2~MHz  measured after arrival in D\"usseldorf (hatched area in Fig.~\ref{fig:drift}) was probably due to a change in the cavity temperature. During the transport the temperature of the aluminum tube enclosing the vacuum system remained unchanged as monitored by the error signal of the temperature control. However at D\"usseldorf the laboratory temperature was above the set point of the temperature control that allowed heating only. This caused an increase in the temperature by about 0.4~K before the problem was discovered and the room temperature was lowered again.
The very large drift rate of 5.3~Hz/s observed during the first day of  measurement could be explained by the cavity cooling down to its set temperature at a rate of $-2$~mK/h. Back in Braunschweig, the original resonance frequency of the cavity was nearly recovered. We observed a reduced drift rate and a small remaining frequency offset, which we attribute to relaxation of stress in the cavity mounting during the transport. Apart from the changed drift rate we did not observe any degradation of the system's performance. Before and after the transport the laser was regularly used to record spectra of the $^{87}$Sr $^1$S$_0$ -- $^3$P$_0$ transition with a Fourier-limited linewidth of 9~Hz and close to 90\% excitation probability\cite{fal11}.
%\sout{ the laboratory temperature temporarily exceeding the range of the cavity's unidirectional temperature control.}}

%\subsection*{Conclusion and outlook}

In summary, we have demonstrated, for the first time, that a high-performance clock laser can be transported without difficulty and set up in a short time for enabling precision metrology applications.
In the near future, the transportable clock laser could be combined with a transportable fiber-based frequency comb for even higher flexibility. 

\section*{Acknowledgment}

The support by the Centre of Quantum Engineering and Space-Time Research (QUEST), 
the European Community's ERA-NET-Plus Programme (Grant No.~217257), 
and by ESA and DLR in the project {\it Space Optical Clocks} is gratefully acknowledged. 
A.N. and S.S. thank T.~Rosenband, S.~Webster, and L.~Lorini for helpful discussions and assistance on cavity design.

% BibTeX users please use
%\bibliography{O:/4-3/4-3-Alle/Papers/TeXBib/TeXBi431}
%\bibliography{texbi431} % ge�ndert Riehle

\begin{thebibliography}{10}

\bibitem{sch09}
S. Schiller {\it et~al.}, Experimental Astronomy {\bf 23},  573  (2009).

\bibitem{wol09}
P. Wolf {\it et~al.}, Experimental Astronomy {\bf 23},  651  (2009).

\bibitem{sch09b}
M. Schioppo {\it et~al.}, Proc. Europ. Time and Frequency Forum 2010.

\bibitem{sch10d}
S. Schiller {\it et~al.}, Proc. Intl. Conf. Space Optics (ESA, 2010).

\bibitem{lei11}
D.~R. Leibrandt {\it et~al.}, Opt. Express {\bf 19},  3471  (2011).

\bibitem{not05}
M. Notcutt, L.-S. Ma, J. Ye, and J.~L. Hall, Opt. Lett. {\bf 30},  1815
  (2005).

\bibitem{naz06}
T. Nazarova, F. Riehle, and U. Sterr, Appl. Phys.~B {\bf 83},  531  (2006).

\bibitem{web07}
S.~A. Webster, M. Oxborrow, and P. Gill, Phys. Rev.~A {\bf 75},  011801(R)
  (2007).

\bibitem{leg09}
T. Legero {\it et~al.}, IEEE Trans. Instrum. Meas. {\bf 58},  1252  (2009).

\bibitem{ma94}
L.-S. Ma, P. Jungner, J. Ye, and J.~L. Hall, Opt. Lett. {\bf 19},  1777
  (1994).

\bibitem{wil08a}
P.~A. Williams, W.~C. Swann, and N.~R. Newbury, J. Opt. Soc. Am.~B {\bf 25},
  1284  (2008).

\bibitem{Nev08}
A.~Y. Nevsky {\it et~al.}, Appl. Phys.~B {\bf 92},  501  (2008).

\bibitem{tel02b}
H.~R. Telle, B. Lipphardt, and J. Stenger, Appl. Phys.~B {\bf 74},  1  (2002).

\bibitem{Shm06}
Y. Shmaliy, {\it Continuous-Time Signals, Signals and Communication Technology}
  (Springer, 2006).

\bibitem{fal11}
S. Falke {\it et~al.}, arXiv:1104.4850v1 [physics.atom-ph], submitted to Metrologia (2011).

\end{thebibliography}
%\bibliographystyle{O:/4-3/4-3-Alle/Papers/TeXBib/prsty}
%\bibliographystyle{Springer}

\end{document}